\documentclass[%
 reprint,
nofootinbib,
 amsmath,amssymb,
 aps,
longbibliography
]{revtex4-1}

\usepackage{graphicx}
\usepackage{dcolumn}
\usepackage{bm}
\usepackage{hyperref}
\usepackage[none]{hyphenat}

\begin{document}

\title{Delayed coincidence in electron-neutrino capture on gallium for neutrino spectroscopy}

\author{Zhe Wang}
\email{Correspondence: wangzhe-hep@mail.tsinghua.edu.cn}
\author{Benda Xu}
\author{Shaomin Chen}
\affiliation{Department~of~Engineering~Physics, Tsinghua~University, Beijing 100084, China}
\affiliation{Key Laboratory of Particle \& Radiation Imaging (Tsinghua University), Ministry of Education, Beijing 100084, China}
\affiliation{Center for High Energy Physics, Tsinghua~University, Beijing 100084, China}

\date{\today}

\begin{abstract}
This work explains a delayed-coincidence method to perform MeV-scale neutrino spectroscopy with electron-neutrino capture on gallium.
An electron-neutrino possessing energy greater than 407.6 keV can be captured on gallium and produce a daughter germanium nucleus at its first excited state with a mean lifetime of 114 ns. The released electron and gammas before the first excited state of Ge can generate a prompt signal representing the solar neutrino energy and the gamma from the deexcitation of the first excited state, 175 keV, can give a delayed signal. The cross-section of this electron-neutrino capture process is evaluated and is comparable with electron-neutrino capture on chlorine. A possible implementation with a liquid scintillator is discussed to exploit the delayed coincidence.
The detection scheme is more feasible than using $^{115}$In, etc., but need a larger target.
The proposed method can be helpful for the MeV-scale solar neutrino spectroscopy and for solving the gallium anomaly.
\end{abstract}

\maketitle

\section{Introduction}
The early works by J.~Bahcall~\cite{JohnWeb} and R.~Davis~\cite{Cleveland:1998nv}, and many results achieved later by other experiments such as
GALLEX~\cite{Hampel:1998xg}, GNO~\cite{Altmann:2005ix}, SAGE~\cite{Abdurashitov:2002nt}, SNO~\cite{Aharmim:2011vm}, Super Kamiokande~\cite{Abe:2016nxk}, KamLAND~\cite{Abe:2011em}, and Borexino~\cite{Agostini:2017ixy}
mark the legend of modern studies of solar neutrino. The standard solar model and neutrino oscillations are the most noteworthy achievements.

However, several theoretical and experimental puzzles continue to exist.
The ``upturn'' effect, which is the increase in solar electron-neutrino survival probability
from matter dominant region to vacuum region, is still poorly constrained by experiments
~\cite{PhysRevD.69.113002, Bonventre:2013wia, Friedland:2004pp, Maltoni:2015kca}.
The carbon-nitrogen-oxygen (CNO) fusion neutrinos from the Sun, which are dominant of the fueling processes of high-temperature massive stars,
have not been observed~\cite{Agostini:2017ixy}\footnote{After the submission of this work, the Borexino collaboration reported a measurement of the CNO neutrino flux of $(7.0^{+3.0}_{-2.0})\times10^8~\rm{cm}^{-2}~\rm{s}^{-1}$ at the ICHEP 2020 conference.}.
For the metal-element abundance prediction of the Sun, a conflict exists
between the helioseismic prediction and the standard-solar-model assumption~\cite{Haxton:2014vqa, Serenelli:2009yc, Serenelli:2011py}.
Moreover, in the source tests with $^{51}$Cr and $^{37}$Ar at the GALLAX and SAGE experiments,
the neutrino flux observations show a deficit from the predictions~\cite{1998114, PhysRevC.59.2246}, known as ``gallium anomaly''.
Experimental and theoretical progress in the future is expected to address these issues.

Electron-neutrino capture on a nucleus,
\begin{eqnarray}
\nu_e + (A, Z) &\rightarrow& (A, Z+1) + e^-,
\label{eq:prompt}
\end{eqnarray}
provides a detection channel of the pure weak charged-current (CC) interaction for MeV-scale neutrino spectroscopy, in which
the energies of the neutrino and the emitted electron simply differ by an energy threshold~\cite{Bahcall:1964zk}.
Neutrino-electron scattering also provides a detection channel of the charged current together with the neutral-current interaction,
but the energy of the recoiled electron is a complicated function of the neutrino energy and the electron emission angle, for example, as shown in~\cite{book}.
Comparing these two types of detection channels, a large number of target electrons are easier to collect in water or an organic liquid scintillator,
but the $\nu_e$-nucleus capture is more convenient to extract the physics depending on the neutrino energy, such as the upturn effect and the structure of CNO neutrinos.

The SNO experiment measured the $e^-$ kinetic energy spectrum above 3.5 MeV~\cite{Aharmim:2011vm} using the CC process $\nu_e+d \rightarrow p+p+e^-$ which has a reaction threshold 1.44 MeV.
All the rest $\nu_e$-nucleus capture measurements have only chemically detected the final state nuclei and no energy measurement of the emitted electrons is available so far.
None of them has effectively addressed the upturn feature and CNO neutrinos.

The delayed-coincidence technique for $\nu_e$-nucleus capture has been actively discussed for nuclei, such as $^{115}$In, $^{100}$Mo, $^{176}$Yb, $^{116}$Cd, etc.~\cite{Raghavan:1976yc, Ejiri:1999rk, Raghavan:1997ad, Zuber:2002wi} because it can greatly decrease the demand for the radioactive purity of the detection material and distinguish the signal from the background $\nu$-electron scattering.
The experimental efforts are still in progress, and facing a lot of technical challenges~\cite{YEH2007329}, for example, the high radioactivity of $^{115}$In.

\section{Delayed coincidence in $\nu_e$ capture on gallium}
\begin{figure}[h]
\includegraphics[width=0.40\textwidth]{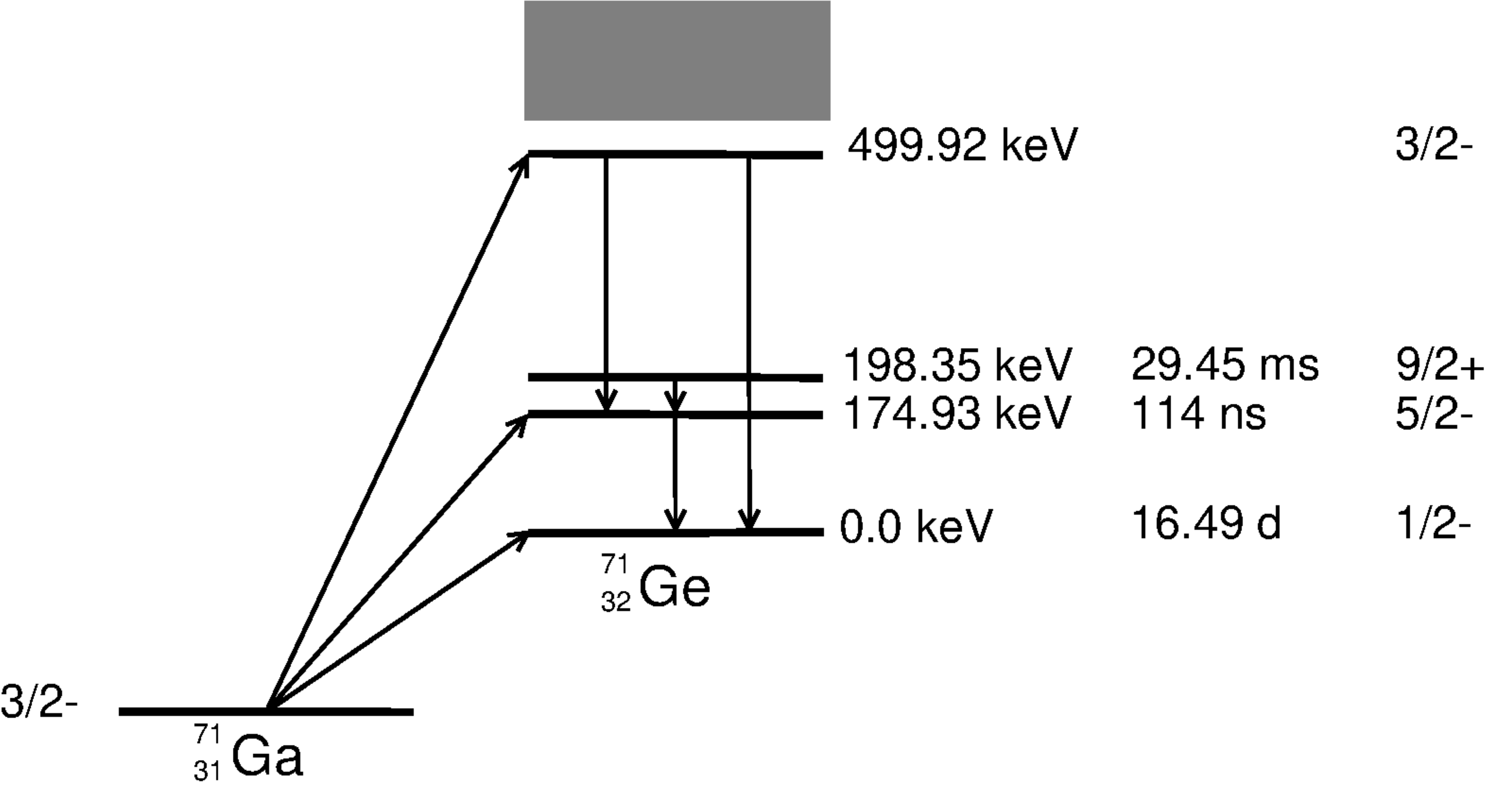} 
\caption{\label{fig:levels} The relevant energy levels of $^{71}\rm{Ga}$-$^{71}\rm{Ge}$ system. The energy level, spin, and parity are labeled. The mean lifetime of the ground, first, and second excited states of $^{71}\rm{Ge}$ are also labeled.
The deexcitation of the first excited state of $^{71}\rm{Ge}$ can be used as a delayed coincidence in the experiment.}
\end{figure}

In this work, we propose a delayed-coincidence method to use gallium to perform a study on solar neutrino spectroscopy.
An energy level plot of the Ga-Ge system is presented in Fig.~\ref{fig:levels.pdf} for a few relevant states.
The process is described in detail below,
\begin{eqnarray}
\nu_e + ^{71}_{31}\rm{Ga} &\rightarrow& ^{71}_{32}\rm{Ge}_{ex} + e^- \rightarrow ^{71}_{32}\rm{Ge}_{1st} + e^- + (\gamma's),
\label{eq:Ga}
\end{eqnarray}
which is followed by
\begin{eqnarray}
^{71}_{32}\rm{Ge}_{1st} \rightarrow ^{71}_{32}\rm{Ge}_{gs} + \gamma~(\rm{174.94~keV}),~\tau=114~ns.
\label{eq:delayed}
\end{eqnarray}
Through the charge-current reaction an electron-neutrino is captured on $^{71}\rm{Ga}$ and emits an electron and a $^{71}\rm{Ge}$ nucleus at an excited state.
The direct nucleus product can be the first excited state of $^{71}\rm{Ge}$ at 175 keV (5/2-) or other higher levels,
which can subsequently decay to the first excited state through one or a few gamma decays with certain probabilities.
The first excited state has a mean lifetime of 114 ns~\cite{PhysRev.97.1033, MORGENSTERN1968370, TAFF1978189}.
The electron and deexcitation gammas in Eq.~\ref{eq:prompt} form a prompt signal.
The neutrino energy, $E_{\nu}$, is the sum of the prompt energy, $E_{\rm{prompt}}$, and the reaction threshold~\cite{ensdf, PhysRevC.56.3391},
\begin{eqnarray}
E_{\nu} = E_{\rm{prompt}} + 407.63~\rm{keV}.
\label{eq:E_prompt}
\end{eqnarray}
The deexcitation of the first excited state in Eq.~\ref{eq:delayed} results in a delayed signal and coincidence.
Note that there is a level at 198 keV with a lifetime of 29.45 ms and must be taken into consideration (Fig.~\ref{fig:levels}).
The gallium can be dissolved in a liquid scintillator and the delay-coincidence signals can be detected with
a photomultiplier tube (PMT) array and a modern fast electronic readout system.

\section{Cross-section}
The differential capture cross-sections of each $^{71}\rm{Ge}$ energy level, $l$, can be calculated with the approach described in~\cite{BahcallBook, RevModPhys.50.881, Giunti:2012tn, Barinov:2017ymq},
\begin{eqnarray}
\label{eq:1}
\sigma_{l}(\omega_e) = \sigma_0 \frac{B({\rm{GT}})_{l}}{B({\rm{GT}})_{\rm{gs}}} \frac{\omega_e p_e}{2\pi\alpha Z} F(Z,\omega_e),
\end{eqnarray}
where $B({\rm{GT}})$s are the Gamow-Teller strengths for each level,
$\sigma_0$ was introduced in~\cite{BahcallBook, RevModPhys.50.881} in which the ground state transition strength was considered,
$\omega_e$ and $p_e$ are the energy and momentum of the electron in the unit of electron mass, respectively,
$\alpha$ is the fine-structure constant, and
$F$ is the Fermi function with $Z=32$ and $\omega_e$, which can be calculated according to~\cite{BahcallBook, MattWeisskopf}.
The $B(\rm{GT})$ values were measured through the $^{71}\rm{Ga}(^3\rm{He},t )^{71}\rm{Ge}$ charge-exchange reaction in~\cite{PhysRevC.91.034608}
and are listed in Table~\ref{tab:BGT}.
For a conservative cross-section estimation, we didn't take the shell-model calculation result for B(GT) of the 5/2- level, which is five times larger~\cite{Haxton:1998uc}.

To evaluate the total cross section, $\sigma$, the incident neutrino energy spectrum $\phi(E_\nu)$, all the allowed excited energy levels,
and their branching ratios to the first excited state, ${\rm{BR}}_l$, are considered
\begin{eqnarray}
\label{eq:2}
\sigma = \int_{E_{\rm{min}}}^{E_{\rm{max}}} \left[ \sum_l \sigma_{l}(\omega_e) {\rm{BR}}_l \right] \phi(E_{\nu}) dE_{\nu},
\end{eqnarray}
where the energy integral region [$E_{\rm{min}}$, $E_{\rm{max}}$] is limited by the $E_{\rm{prompt}}$ of interest for detection
and the contributions through the 198 keV level are excluded.
The branching ratios of the excited states to the first excited state
can be calculated based on the information given in ENSDF~\cite{ensdf}.
However, only the data up to approximately 2 MeV are available for this estimation.
Table~\ref{tab:BGT} summarizes the branching fractions.

With Eq.~\ref{eq:1} and~\ref{eq:2} and the input in Table~\ref{tab:BGT},
the capture cross-sections of solar pp, $^7$Be, pep, and CNO neutrinos and $^{51}$Cr and $^{37}$Ar neutrinos are calculated and listed in Table~\ref{tab:SolarX} where the solar neutrino spectrum input is from~\cite{JohnWeb} and the $^{51}$Cr and $^{37}$Ar spectrum input is from~\cite{PhysRevC.59.2246}.

\begin{table}[]
\caption{\label{tab:BGT}%
$^{71}\rm{Ge}$ energy levels for $\nu_e$ capture on $^{71}\rm{Ga}$ with the non-zero measured $B(\rm{GT})$ values, and the branching fractions to the first excited state at 175 keV.}
\begin{ruledtabular}
\begin{tabular}{ccc}
$^{71}_{32}\rm{Ge}$ levels & $B$(GT)           & BR   \\
                  MeV        & $\times10^{-2}$ &      \\
\colrule
0	    & 8.52(11)& 0          \\
174.9	& 0.34(26)& 1          \\
499.9	& 1.76(14)& 0.004777   \\
708.2	& 0.11(5) & 0.04489    \\
808.2	& 2.29(10)& 0.241717   \\
1095.5	& 1.83(17)&	0.0725541  \\
1298.7	& 1.33(8) &	0.0249493  \\
1378.6	& 0.33(4) &	0.409911   \\
1598.5	& 0.11(5) &	0.201239   \\
1743.4	& 0.68(2) &	0.170329   \\
1965    & 0.12(6) & 0          \\
\end{tabular}
\end{ruledtabular}
\end{table}

\begin{table}[]
\caption{\label{tab:SolarX}%
Cross-sections for the solar pp, $^7$Be, pep, and CNO neutrinos and $^{51}$Cr and $^{37}$Ar neutrinos.
For each neutrino component either the maximum energies for continuous distributions
or the peak energies for the discrete structures are listed in column two.
Columns three and four list their cross-sections and the ones with $E_{\rm{prompt}}$ beyond 0.2 MeV, respectively.
For the $^7$Be neutrinos, results are given for the two separate energy components at 380 and 860 keV.
For the $^{51}$Cr and $^{37}$Ar neutrinos, results are given for all neutrino lines and summed with their corresponding branching ratios~\cite{PhysRevC.59.2246}.
}
\begin{ruledtabular}
\begin{tabular}{cccc}
           & $E$  &   $\sigma$             &  $\sigma~(>0.2 MeV)$     \\
           & MeV  &   $10^{-46}~\rm{cm}^2$ &  $10^{-46}~\rm{cm}^2$    \\
\colrule
pp       & $<$0.42 MeV & $6.6\times10^{-3}$  & 0 \\
pep      & 1.45 MeV    & 8.2  & 8.2 \\
$^7$Be-380  & 0.38 MeV    & 0    & 0 \\
$^7$Be-860  & 0.86 MeV    & 1.96 & 1.96 \\
$^{13}$N & $<$1.19 MeV & 1.52 & 1.37 \\
$^{15}$O & $<$1.73 MeV & 3.87 & 3.79 \\
$^{17}$F & $<$1.74 MeV & 3.90 & 3.83 \\
\colrule
$^{51}$Cr& 0.747 MeV   & 1.49 & 1.49 \\
$^{51}$Cr& 0.752 MeV   & 1.51 & 1.51 \\
$^{51}$Cr& 0.427 MeV   & 0.51 & 0 \\
$^{51}$Cr& 0.432 MeV   & 0.52 & 0 \\
$^{51}$Cr& all         & 1.39 & 1.34 \\
\colrule
$^{37}$Ar& 0.811 MeV   & 1.74 & 1.74 \\
$^{37}$Ar& 0.813 MeV   & 1.74 & 1.74 \\
$^{37}$Ar& all         & 1.74 & 1.74 \\
\end{tabular}
\end{ruledtabular}
\end{table}

\section{Gallium-loaded liquid scintillator detector}
One measurement scheme is a Ga-loaded liquid scintillator.
The Ga-loaded liquid scintillator is contained in a spherical transparent container.
The scintillation and Cherenkov lights emitted by the prompt and delayed signals are detected by PMTs,
and subsequently read out with fast electronics.

There are several applications to load different elements into an organic liquid scintillator.
The Te-loaded liquid scintillator~\cite{BILLER2015205}, In-loaded liquid scintillator~\cite{Raghavan_2008},
Gd-loaded liquid scintillator~\cite{YEH2007329}, and Li-loaded liquid scintillator~\cite{Ashenfelter:2018cli} are some examples.
A method to load other elements into water-based liquid scintillators also exists~\cite{YEH201151}.

The timing feature of the PMT and electronics is suitable for such a method.
The rising time of the large diameter PMTs ($\ge$8 inches)~\cite{Hamamatsu} is less than 10 ns and their transition time spread is 2-4 ns.
Fast waveform digitizers can be found~\cite{caen} with a sampling rate of 1 Giga samples per second and no dead time.

The random coincident background from natural radioactivities is negligible for a 342 ns coincident window (3$\times$lifetime) according to the calculation method in ~\cite{Raghavan:1976yc, Ejiri:1999rk}.

The late pulse of the PMTs should be considered.
Actual PMT time residual distributions in a large detector are presented by the SNO experiment (Fig. 5 of~\cite{Aharmim:2005gt} and Fig. 3 of~\cite{Aharmim:2006kv}),
where the late pulse rate is approximately 1\% of the major pulse height per 0.33 ns from 10 to 80 ns because of the PMT's, electronics and reflections, and only 0.2\% per 0.33 ns after 80 ns.
Considering a light yield of 500 photoelectrons (PE) per MeV in a liquid scintillator detector and a prompt 2 MeV signal, the number of PE for the prompt and delayed signals are 1000 and 88, respectively. After 80 ns, the number of PE for the late pulse in a 20 ns window is approximately 4,
so that identifying the 175-keV delayed signal is not a concern.

\section{Requirement for CNO neutrinos}
The requirement for the CNO neutrino detection will be discussed in this session.
Both statistics and detector energy resolution are important for CNO neutrino detection and for separating each CNO component.

The statistics of the neutrino signals, $N$, can be calculated with the neutrino flux as
\begin{eqnarray}
\label{eq:solarrate}
N = flux\times\sigma\times N_{\rm{Ga71}}\times t \times p,
\end{eqnarray}
where $N_{\rm{Ga71}}$ is the number of target nuclei, $t$ is the data-taking time, and $p$ is the survival probability.
So the product of target Ga mass, $m$, and data-taking time, $t$, is the first key factor to consider.
With the CNO neutrino flux prediction given in the low-metallicity solar model~\cite{AGS09} and the neutrino oscillation parameters~\cite{PDG2018},
the event rates of $^{13}$N and $^{15}$O+$^{17}$F are
\begin{eqnarray}
\label{eq:Rate}
\rm{Rate}(^{13}{N})= 100/(52.9~kton~year),\\
\rm{Rate}(^{15}{O}+^{17}{F})= 100/(29.4~kton~year),
\end{eqnarray}
where the mass is for natural gallium and the signals of $^{15}$O and $^{17}$F are treated together, because their spectra are indistinguishable and these conventions are used for the whole paper.
We found that 52.9 kton year and 29.4 kton year are the thresholds to reach 10\% precision for $^{13}$N and $^{15}$O+$^{17}$F measurement, respectively.

\begin{figure}[h]
\includegraphics[width=0.48\textwidth]{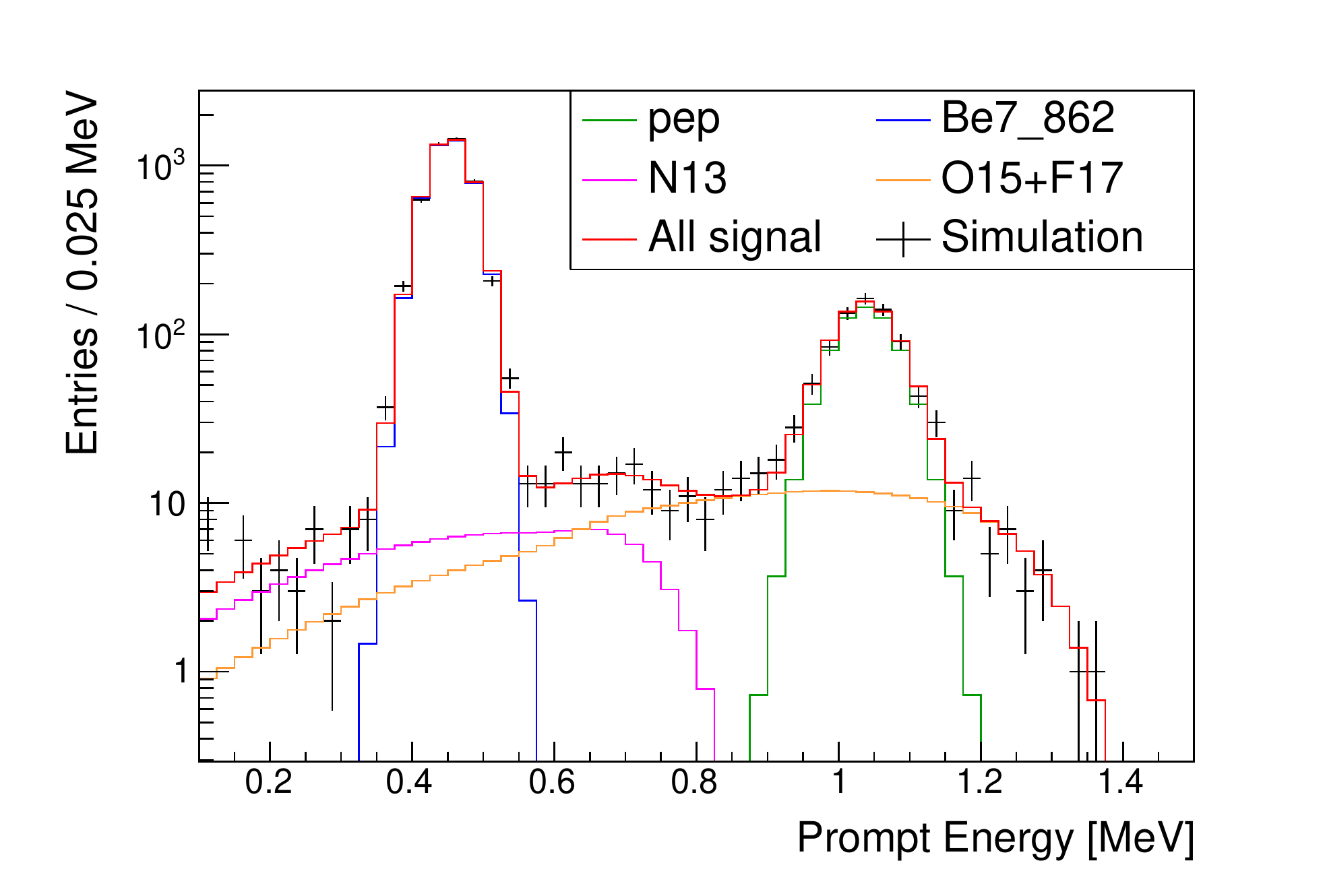} 
\caption{\label{fig:CNO} Solar neutrino simulation and fit results in the CNO neutrino detection region. The product of the target natural Ga mass and data-taking time is set to 100 kton year and the light yield is set to 500 PE/MeV.}
\end{figure}

A good energy resolution, $\sigma_E$, is also expected so the $^{13}$N and $^{15}$O+$^{17}$F neutrinos can be distinguished from other solar neutrino components, i.e.~the $^{7}$Be and pep neutrinos, and the feature can be seen in Fig.~\ref{fig:CNO}.
The energy resolution is governed by the light yield, $L$ (number of photoelectrons, PE, per MeV) with $\sigma_E\approx1/\sqrt L$.

To give a better understanding of the requirement, we simulated the solar neutrino signals according to the neutrino fluxes predicted by the low-metallicity solar model~\cite{AGS09} and a few detector configurations with different $m\times t$ and $L$. The simulated spectrum is then fitted with the known solar neutrino spectra and the fit result gives the expected uncertainty in determining the $^{13}$N and $^{15}$O+$^{17}$F fluxes.
An example can be seen in Fig.~\ref{fig:CNO}.
Here the detection efficiency is assumed to be 100\% and no other background is considered.
More detail of the method can be found in Ref.~\cite{Jinping}.
After scanning through several detector configurations, Tab.~\ref{tab:CNO} gives a few promising results.
In summary, assuming a 20-year data-taking time, a few kton of natural gallium is needed to measure the $^{13}$N and $^{15}$O+$^{17}$F fluxes.

\begin{table}[h]
\caption{Relative precision, error/flux, in determining solar $^{13}$N and $^{15}$O+$^{17}$F neutrino fluxes
for different target mass $\times$ data-taking time and light yield.
\label{tab:CNO}}
\footnotesize
\begin{ruledtabular}
\begin{tabular}{clcc}
                      &                   & 200 PE/MeV & 500 PE/MeV \\\hline
 25 kton year  & $^{13}$N          & 0.37       & 0.25  \\
                      & $^{15}$O+$^{17}$F & 0.25       & 0.19  \\\hline
 50 kton year  & $^{13}$N          & 0.23       & 0.20  \\
                      & $^{15}$O+$^{17}$F & 0.18       & 0.14  \\\hline
100 kton year  & $^{13}$N          & 0.17       & 0.13  \\
                      & $^{15}$O+$^{17}$F & 0.12       & 0.09  \\
\end{tabular}
\end{ruledtabular}
\end{table}

\section{Requirement for $^{51}$C\lowercase{r} neutrinos}
To study $^{51}$Cr neutrino capture cross-section, the number of $^{51}$Cr signal, N, is estimated as
\begin{eqnarray}
\label{eq:NCr}
N = A\times t \times \rho_{\rm{Ga71}} \times R \times \sigma,
\end{eqnarray}
where $A$ is the strength of the source, $t$ is the data-taking time, $\rho_{\rm{Ga71}}$ is the number density of $^{71}$Ga, and $R$ is the path length through Ga.
We assume an intensive $^{51}$Cr source is placed in the center of a spherical detector and $R$ is the radius of the sensitive region with 100\% detection efficiency.
The source has a constant rate of 50 PBq similar to GALLEX and SAGE~\cite{1998114, PhysRevC.59.2246}.
The Ga-loaded liquid scintillator has a density of 1 g/cm$^3$ and the natural Ga mass fraction is 10\%.
We must say the above assumptions are not realistic at all, but they will provide a useful estimation for the order of magnitude.
The calculation shows
\begin{eqnarray}
\label{eq:RateCr}
\rm{Rate}(^{51}{Cr})= 100/(13~meter~year),\\
\rm{Rate}(^{51}{Cr})= 100/(0.97~kton~year).
\end{eqnarray}
To get at least 100 signals, the product of the detector radius and data-taking time must be larger than 13 meter year, or equivalently the natural Ga mass times data-taking time larger than 0.97 kton year.

\section{Discussion and conclusion}
In this work we explained the delayed-coincidence method in $\nu_e$ capture on $^{71}\rm{Ga}$ and evaluated the cross-sections below 2 MeV.

Naturally, gallium has only two stable isotopes and the natural abundance of $^{71}\rm{Ga}$ is 39.9\%.
This overcomes the difficulty of intrinsic background in $^{115}\rm{In}$~\cite{Raghavan:1976yc} and $^{100}\rm{Mo}$~\cite{Ejiri:1999rk}.
The coincident time of $^{71}\rm{Ga}$ is the shortest among $^{115}$In, $^{100}$Mo, and $^{116}$Cd~\cite{Zuber:2002wi} and is just over the limitation of PMT system.
The delayed energy is higher than $^{176}$Yb~\cite{Raghavan:1997ad} and is also higher than the endpoint of $^{14}$C, which is contained in organic detection material.
Experimentally it is easier to implement.

The cross-section going through the first excited state of $^{71}\rm{Ge}$ is
2\%-3\% of that going through the ground state and much smaller than $^{115}$In; however it is comparable with that of $^{37}\rm{Cl}$~\cite{BahcallBook} which was used by R.~Davis.
If we take the B(BG) for the 5/2- level of shell-model prediction~\cite{Haxton:1998uc}, the cross-section is five times larger.
This is the largest uncertainty and at this moment, we cannot exclude one or the other.
The cross-section beyond 2 MeV can be measured using a neutrino source on a spallation neutron facility~\cite{Bolozdynya:2012xv, CSNS} with a known $\nu_e$ spectrum from the muon decay.

The application of loading Ga into a liquid scintillator is also discussed.
The delayed signal is well detectable by a modern PMT-digitizer array.
The background situation is promising except for some efficiency loss due to the late PMT pulses.

In the search of the CNO neutrinos, this method
is clear of the unfolding complexity of the electron targets and
can be useful for directly measuring the solar $\nu_e$ energy spectrum and
extracting $^{13}$N and $^{15}$O+$^{17}$F neutrinos separately.

For the gallium anomaly, the method could be interesting, because the considerable focus of the cause is on the neutrino
capture cross-section prediction for the first excited state of $^{71}\rm{Ge}$~\cite{Haxton:1998uc, PhysRevC.91.034608, Giunti:2012tn}.
This method will give an insight into the popular speculation on the systematic uncertainty of the first excited state of $^{71}$Ge.

To reach a 10\% precision of $^{13}$N, $^{15}$O+$^{17}$F, and $^{51}$Cr neutrinos, the need for natural Ga mass is at kton scale.

In summary, the delayed-coincidence in $^{71}$Ga($\nu_e$,$e^-$)$^{71}$Ge reaction involving the first excited state of $^{71}$Ge
could be technically applicable for MeV neutrino spectroscopy. For the studies of gallium anomaly, upturn, and CNO neutrinos,
it is an approach to consider.

\section*{Acknowledgement}
This work is supported in part by
the National Natural Science Foundation of China (Nos.~11620101004, 11475093, and 11235006),
the Ministry of Science and Technology of China (No. 2018YFA0404102),
the Key Laboratory of Particle \& Radiation Imaging (Tsinghua University),
and the CAS Center for Excellence in Particle Physics (CCEPP).

\nocite{*}

\bibliography{Gallium}

\begin{thebibliography}{50}%
\makeatletter
\providecommand \@ifxundefined [1]{%
 \@ifx{#1\undefined}
}%
\providecommand \@ifnum [1]{%
 \ifnum #1\expandafter \@firstoftwo
 \else \expandafter \@secondoftwo
 \fi
}%
\providecommand \@ifx [1]{%
 \ifx #1\expandafter \@firstoftwo
 \else \expandafter \@secondoftwo
 \fi
}%
\providecommand \natexlab [1]{#1}%
\providecommand \enquote  [1]{``#1''}%
\providecommand \bibnamefont  [1]{#1}%
\providecommand \bibfnamefont [1]{#1}%
\providecommand \citenamefont [1]{#1}%
\providecommand \href@noop [0]{\@secondoftwo}%
\providecommand \href [0]{\begingroup \@sanitize@url \@href}%
\providecommand \@href[1]{\@@startlink{#1}\@@href}%
\providecommand \@@href[1]{\endgroup#1\@@endlink}%
\providecommand \@sanitize@url [0]{\catcode `\\12\catcode `\$12\catcode
  `\&12\catcode `\#12\catcode `\^12\catcode `\_12\catcode `\%12\relax}%
\providecommand \@@startlink[1]{}%
\providecommand \@@endlink[0]{}%
\providecommand \url  [0]{\begingroup\@sanitize@url \@url }%
\providecommand \@url [1]{\endgroup\@href {#1}{\urlprefix }}%
\providecommand \urlprefix  [0]{URL }%
\providecommand \Eprint [0]{\href }%
\providecommand \doibase [0]{http://dx.doi.org/}%
\providecommand \selectlanguage [0]{\@gobble}%
\providecommand \bibinfo  [0]{\@secondoftwo}%
\providecommand \bibfield  [0]{\@secondoftwo}%
\providecommand \translation [1]{[#1]}%
\providecommand \BibitemOpen [0]{}%
\providecommand \bibitemStop [0]{}%
\providecommand \bibitemNoStop [0]{.\EOS\space}%
\providecommand \EOS [0]{\spacefactor3000\relax}%
\providecommand \BibitemShut  [1]{\csname bibitem#1\endcsname}%
\let\auto@bib@innerbib\@empty
\bibitem [{Joh()}]{JohnWeb}%
  \BibitemOpen
  \href@noop {} {}\bibinfo {note} {J.~Bahcall home page:
  http://www.sns.ias.edu/\~{}jnb/}\BibitemShut {NoStop}%
\bibitem [{\citenamefont {Cleveland}\ \emph {et~al.}(1998)\citenamefont
  {Cleveland} \emph {et~al.}}]{Cleveland:1998nv}%
  \BibitemOpen
  \bibfield  {author} {\bibinfo {author} {\bibfnamefont {B.~T.}\ \bibnamefont
  {Cleveland}} \emph {et~al.},\ }\bibfield  {title} {\enquote {\bibinfo {title}
  {{Measurement of the solar electron neutrino flux with the Homestake chlorine
  detector}},}\ }\href {\doibase 10.1086/305343} {\bibfield  {journal}
  {\bibinfo  {journal} {Astrophys. J.}\ }\textbf {\bibinfo {volume} {496}},\
  \bibinfo {pages} {505--526} (\bibinfo {year} {1998})}\BibitemShut {NoStop}%
\bibitem [{\citenamefont {Hampel}\ \emph {et~al.}(1999)\citenamefont {Hampel}
  \emph {et~al.}}]{Hampel:1998xg}%
  \BibitemOpen
  \bibfield  {author} {\bibinfo {author} {\bibfnamefont {W.}~\bibnamefont
  {Hampel}} \emph {et~al.} (\bibinfo {collaboration} {GALLEX}),\ }\bibfield
  {title} {\enquote {\bibinfo {title} {{GALLEX solar neutrino observations:
  Results for GALLEX IV}},}\ }\href {\doibase 10.1016/S0370-2693(98)01579-2}
  {\bibfield  {journal} {\bibinfo  {journal} {Phys. Lett.}\ }\textbf {\bibinfo
  {volume} {B447}},\ \bibinfo {pages} {127--133} (\bibinfo {year}
  {1999})}\BibitemShut {NoStop}%
\bibitem [{\citenamefont {Altmann}\ \emph {et~al.}(2005)\citenamefont {Altmann}
  \emph {et~al.}}]{Altmann:2005ix}%
  \BibitemOpen
  \bibfield  {author} {\bibinfo {author} {\bibfnamefont {M.}~\bibnamefont
  {Altmann}} \emph {et~al.} (\bibinfo {collaboration} {GNO}),\ }\bibfield
  {title} {\enquote {\bibinfo {title} {{Complete results for five years of GNO
  solar neutrino observations}},}\ }\href {\doibase
  10.1016/j.physletb.2005.04.068} {\bibfield  {journal} {\bibinfo  {journal}
  {Phys. Lett.}\ }\textbf {\bibinfo {volume} {B616}},\ \bibinfo {pages}
  {174--190} (\bibinfo {year} {2005})}\BibitemShut {NoStop}%
\bibitem [{\citenamefont {Abdurashitov}\ \emph {et~al.}(2002)\citenamefont
  {Abdurashitov} \emph {et~al.}}]{Abdurashitov:2002nt}%
  \BibitemOpen
  \bibfield  {author} {\bibinfo {author} {\bibfnamefont {J.~N.}\ \bibnamefont
  {Abdurashitov}} \emph {et~al.} (\bibinfo {collaboration} {SAGE}),\ }\bibfield
   {title} {\enquote {\bibinfo {title} {{Solar neutrino flux measurements by
  the Soviet-American Gallium Experiment (SAGE) for half the 22 year solar
  cycle}},}\ }\href {\doibase 10.1134/1.1506424} {\bibfield  {journal}
  {\bibinfo  {journal} {J. Exp. Theor. Phys.}\ }\textbf {\bibinfo {volume}
  {95}},\ \bibinfo {pages} {181--193} (\bibinfo {year} {2002})},\ \bibinfo
  {note} {[Zh. Eksp. Teor. Fiz.122,211(2002)]}\BibitemShut {NoStop}%
\bibitem [{\citenamefont {Aharmim}\ \emph {et~al.}(2013)\citenamefont {Aharmim}
  \emph {et~al.}}]{Aharmim:2011vm}%
  \BibitemOpen
  \bibfield  {author} {\bibinfo {author} {\bibfnamefont {B.}~\bibnamefont
  {Aharmim}} \emph {et~al.} (\bibinfo {collaboration} {SNO}),\ }\bibfield
  {title} {\enquote {\bibinfo {title} {{Combined Analysis of all Three Phases
  of Solar Neutrino Data from the Sudbury Neutrino Observatory}},}\ }\href
  {\doibase 10.1103/PhysRevC.88.025501} {\bibfield  {journal} {\bibinfo
  {journal} {Phys. Rev.}\ }\textbf {\bibinfo {volume} {C88}},\ \bibinfo {pages}
  {025501} (\bibinfo {year} {2013})}\BibitemShut {NoStop}%
\bibitem [{\citenamefont {Abe}\ \emph {et~al.}(2016)\citenamefont {Abe} \emph
  {et~al.}}]{Abe:2016nxk}%
  \BibitemOpen
  \bibfield  {author} {\bibinfo {author} {\bibfnamefont {K.}~\bibnamefont
  {Abe}} \emph {et~al.} (\bibinfo {collaboration} {Super-Kamiokande}),\
  }\bibfield  {title} {\enquote {\bibinfo {title} {{Solar Neutrino Measurements
  in Super-Kamiokande-IV}},}\ }\href {\doibase 10.1103/PhysRevD.94.052010}
  {\bibfield  {journal} {\bibinfo  {journal} {Phys. Rev.}\ }\textbf {\bibinfo
  {volume} {D94}},\ \bibinfo {pages} {052010} (\bibinfo {year}
  {2016})}\BibitemShut {NoStop}%
\bibitem [{\citenamefont {Abe}\ \emph {et~al.}(2011)\citenamefont {Abe} \emph
  {et~al.}}]{Abe:2011em}%
  \BibitemOpen
  \bibfield  {author} {\bibinfo {author} {\bibfnamefont {S.}~\bibnamefont
  {Abe}} \emph {et~al.} (\bibinfo {collaboration} {KamLAND}),\ }\bibfield
  {title} {\enquote {\bibinfo {title} {{Measurement of the 8B Solar Neutrino
  Flux with the KamLAND Liquid Scintillator Detector}},}\ }\href {\doibase
  10.1103/PhysRevC.84.035804} {\bibfield  {journal} {\bibinfo  {journal} {Phys.
  Rev.}\ }\textbf {\bibinfo {volume} {C84}},\ \bibinfo {pages} {035804}
  (\bibinfo {year} {2011})}\BibitemShut {NoStop}%
\bibitem [{\citenamefont {Agostini}\ \emph {et~al.}(2019)\citenamefont
  {Agostini} \emph {et~al.}}]{Agostini:2017ixy}%
  \BibitemOpen
  \bibfield  {author} {\bibinfo {author} {\bibfnamefont {M.}~\bibnamefont
  {Agostini}} \emph {et~al.} (\bibinfo {collaboration} {Borexino}),\ }\bibfield
   {title} {\enquote {\bibinfo {title} {{First Simultaneous Precision
  Spectroscopy of $pp$, $^7$Be, and $pep$ Solar Neutrinos with Borexino
  Phase-II}},}\ }\href {\doibase 10.1103/PhysRevD.100.082004} {\bibfield
  {journal} {\bibinfo  {journal} {Phys. Rev.}\ }\textbf {\bibinfo {volume}
  {D100}},\ \bibinfo {pages} {082004} (\bibinfo {year} {2019})}\BibitemShut
  {NoStop}%
\bibitem [{\citenamefont {de~Holanda}\ and\ \citenamefont
  {Smirnov}(2004)}]{PhysRevD.69.113002}%
  \BibitemOpen
  \bibfield  {author} {\bibinfo {author} {\bibfnamefont {P.~C.}\ \bibnamefont
  {de~Holanda}}\ and\ \bibinfo {author} {\bibfnamefont {A.~Yu.}\ \bibnamefont
  {Smirnov}},\ }\bibfield  {title} {\enquote {\bibinfo {title} {Homestake
  result, sterile neutrinos, and low energy solar neutrino experiments},}\
  }\href {\doibase 10.1103/PhysRevD.69.113002} {\bibfield  {journal} {\bibinfo
  {journal} {Phys. Rev. D}\ }\textbf {\bibinfo {volume} {69}},\ \bibinfo
  {pages} {113002} (\bibinfo {year} {2004})}\BibitemShut {NoStop}%
\bibitem [{\citenamefont {Bonventre}\ \emph {et~al.}(2013)\citenamefont
  {Bonventre} \emph {et~al.}}]{Bonventre:2013wia}%
  \BibitemOpen
  \bibfield  {author} {\bibinfo {author} {\bibfnamefont {R.}~\bibnamefont
  {Bonventre}} \emph {et~al.},\ }\bibfield  {title} {\enquote {\bibinfo {title}
  {{Non-Standard Models, Solar Neutrinos, and Large $\theta_{13}$}},}\ }\href
  {\doibase 10.1103/PhysRevD.88.053010} {\bibfield  {journal} {\bibinfo
  {journal} {Phys. Rev.}\ }\textbf {\bibinfo {volume} {D88}},\ \bibinfo {pages}
  {053010} (\bibinfo {year} {2013})}\BibitemShut {NoStop}%
\bibitem [{\citenamefont {Friedland}\ \emph {et~al.}(2004)\citenamefont
  {Friedland}, \citenamefont {Lunardini},\ and\ \citenamefont
  {Pena-Garay}}]{Friedland:2004pp}%
  \BibitemOpen
  \bibfield  {author} {\bibinfo {author} {\bibfnamefont {Alexander}\
  \bibnamefont {Friedland}}, \bibinfo {author} {\bibfnamefont {Cecilia}\
  \bibnamefont {Lunardini}}, \ and\ \bibinfo {author} {\bibfnamefont {Carlos}\
  \bibnamefont {Pena-Garay}},\ }\bibfield  {title} {\enquote {\bibinfo {title}
  {{Solar neutrinos as probes of neutrino matter interactions}},}\ }\href
  {\doibase 10.1016/j.physletb.2004.05.047} {\bibfield  {journal} {\bibinfo
  {journal} {Phys. Lett.}\ }\textbf {\bibinfo {volume} {B594}},\ \bibinfo
  {pages} {347} (\bibinfo {year} {2004})}\BibitemShut {NoStop}%
\bibitem [{\citenamefont {Maltoni}\ and\ \citenamefont
  {Smirnov}(2016)}]{Maltoni:2015kca}%
  \BibitemOpen
  \bibfield  {author} {\bibinfo {author} {\bibfnamefont {Michele}\ \bibnamefont
  {Maltoni}}\ and\ \bibinfo {author} {\bibfnamefont {Alexei~{\relax Yu}.}\
  \bibnamefont {Smirnov}},\ }\bibfield  {title} {\enquote {\bibinfo {title}
  {{Solar neutrinos and neutrino physics}},}\ }\href {\doibase
  10.1140/epja/i2016-16087-0} {\bibfield  {journal} {\bibinfo  {journal} {Eur.
  Phys. J.}\ }\textbf {\bibinfo {volume} {A52}},\ \bibinfo {pages} {87}
  (\bibinfo {year} {2016})}\BibitemShut {NoStop}%
\bibitem [{\citenamefont {Haxton}(2014)}]{Haxton:2014vqa}%
  \BibitemOpen
  \bibfield  {author} {\bibinfo {author} {\bibfnamefont {Wick}\ \bibnamefont
  {Haxton}},\ }\bibfield  {title} {\enquote {\bibinfo {title} {{Neutrino
  physics: What makes the Sun shine}},}\ }\href {\doibase 10.1038/512378a}
  {\bibfield  {journal} {\bibinfo  {journal} {Nature}\ }\textbf {\bibinfo
  {volume} {512}},\ \bibinfo {pages} {378--380} (\bibinfo {year}
  {2014})}\BibitemShut {NoStop}%
\bibitem [{\citenamefont {Serenelli}\ \emph
  {et~al.}(2009{\natexlab{a}})\citenamefont {Serenelli}, \citenamefont {Basu},
  \citenamefont {Ferguson},\ and\ \citenamefont {Asplund}}]{Serenelli:2009yc}%
  \BibitemOpen
  \bibfield  {author} {\bibinfo {author} {\bibfnamefont {Aldo}\ \bibnamefont
  {Serenelli}}, \bibinfo {author} {\bibfnamefont {Sarbani}\ \bibnamefont
  {Basu}}, \bibinfo {author} {\bibfnamefont {Jason~W.}\ \bibnamefont
  {Ferguson}}, \ and\ \bibinfo {author} {\bibfnamefont {Martin}\ \bibnamefont
  {Asplund}},\ }\bibfield  {title} {\enquote {\bibinfo {title} {{New Solar
  Composition: The Problem With Solar Models Revisited}},}\ }\href {\doibase
  10.1088/0004-637X/705/2/L123} {\bibfield  {journal} {\bibinfo  {journal}
  {Astrophys. J.}\ }\textbf {\bibinfo {volume} {705}},\ \bibinfo {pages}
  {L123--L127} (\bibinfo {year} {2009}{\natexlab{a}})}\BibitemShut {NoStop}%
\bibitem [{\citenamefont {Serenelli}\ \emph {et~al.}(2011)\citenamefont
  {Serenelli}, \citenamefont {Haxton},\ and\ \citenamefont
  {Pena-Garay}}]{Serenelli:2011py}%
  \BibitemOpen
  \bibfield  {author} {\bibinfo {author} {\bibfnamefont {Aldo~M.}\ \bibnamefont
  {Serenelli}}, \bibinfo {author} {\bibfnamefont {W.~C.}\ \bibnamefont
  {Haxton}}, \ and\ \bibinfo {author} {\bibfnamefont {Carlos}\ \bibnamefont
  {Pena-Garay}},\ }\bibfield  {title} {\enquote {\bibinfo {title} {{Solar
  models with accretion. I. Application to the solar abundance problem}},}\
  }\href {\doibase 10.1088/0004-637X/743/1/24} {\bibfield  {journal} {\bibinfo
  {journal} {Astrophys. J.}\ }\textbf {\bibinfo {volume} {743}},\ \bibinfo
  {pages} {24} (\bibinfo {year} {2011})}\BibitemShut {NoStop}%
\bibitem [{\citenamefont {Hampel}\ \emph {et~al.}(1998)\citenamefont {Hampel}
  \emph {et~al.}}]{1998114}%
  \BibitemOpen
  \bibfield  {author} {\bibinfo {author} {\bibfnamefont {W}~\bibnamefont
  {Hampel}} \emph {et~al.},\ }\bibfield  {title} {\enquote {\bibinfo {title}
  {Final results of the ${}^{51}\mathrm{Cr}$ neutrino source experiments in
  gallex},}\ }\href {\doibase https://doi.org/10.1016/S0370-2693(97)01562-1}
  {\bibfield  {journal} {\bibinfo  {journal} {Physics Letters B}\ }\textbf
  {\bibinfo {volume} {420}},\ \bibinfo {pages} {114 -- 126} (\bibinfo {year}
  {1998})}\BibitemShut {NoStop}%
\bibitem [{\citenamefont {Abdurashitov}\ \emph {et~al.}(1999)\citenamefont
  {Abdurashitov} \emph {et~al.}}]{PhysRevC.59.2246}%
  \BibitemOpen
  \bibfield  {author} {\bibinfo {author} {\bibfnamefont {J.~N.}\ \bibnamefont
  {Abdurashitov}} \emph {et~al.} (\bibinfo {collaboration} {The SAGE
  Collaboration}),\ }\bibfield  {title} {\enquote {\bibinfo {title}
  {Measurement of the response of a gallium metal solar neutrino experiment to
  neutrinos from a ${}^{51}\mathrm{Cr}$ source},}\ }\href {\doibase
  10.1103/PhysRevC.59.2246} {\bibfield  {journal} {\bibinfo  {journal} {Phys.
  Rev. C}\ }\textbf {\bibinfo {volume} {59}},\ \bibinfo {pages} {2246--2263}
  (\bibinfo {year} {1999})}\BibitemShut {NoStop}%
\bibitem [{\citenamefont {Bahcall}(1964)}]{Bahcall:1964zk}%
  \BibitemOpen
  \bibfield  {author} {\bibinfo {author} {\bibfnamefont {John~N.}\ \bibnamefont
  {Bahcall}},\ }\bibfield  {title} {\enquote {\bibinfo {title} {{Solar neutrino
  cross sections and nuclear beta decay}},}\ }\href {\doibase
  10.1103/PhysRev.135.B137} {\bibfield  {journal} {\bibinfo  {journal} {Phys.
  Rev.}\ }\textbf {\bibinfo {volume} {135}},\ \bibinfo {pages} {B137--B146}
  (\bibinfo {year} {1964})}\BibitemShut {NoStop}%
\bibitem [{\citenamefont {Giunti}\ and\ \citenamefont {Kim}(2007)}]{book}%
  \BibitemOpen
  \bibfield  {author} {\bibinfo {author} {\bibfnamefont {C.}~\bibnamefont
  {Giunti}}\ and\ \bibinfo {author} {\bibfnamefont {C.~W.}\ \bibnamefont
  {Kim}},\ }\href@noop {} {\emph {\bibinfo {title} {Fundamentals of Neutrino
  Physics and Astrophysics}}}\ (\bibinfo  {publisher} {Oxford University
  Press},\ \bibinfo {year} {2007})\BibitemShut {NoStop}%
\bibitem [{\citenamefont {Raghavan}(1976)}]{Raghavan:1976yc}%
  \BibitemOpen
  \bibfield  {author} {\bibinfo {author} {\bibfnamefont {R.~S.}\ \bibnamefont
  {Raghavan}},\ }\bibfield  {title} {\enquote {\bibinfo {title} {{Inverse beta
  decay of 115-In to 115-Sn*: a new possibility for detecting solar neutrinos
  from the proton-proton reaction}},}\ }\href {\doibase
  10.1103/PhysRevLett.37.259} {\bibfield  {journal} {\bibinfo  {journal} {Phys.
  Rev. Lett.}\ }\textbf {\bibinfo {volume} {37}},\ \bibinfo {pages} {259--262}
  (\bibinfo {year} {1976})}\BibitemShut {NoStop}%
\bibitem [{\citenamefont {Ejiri}\ \emph {et~al.}(2000)\citenamefont {Ejiri},
  \citenamefont {Engel}, \citenamefont {Hazama}, \citenamefont {Krastev},
  \citenamefont {Kudomi},\ and\ \citenamefont {Robertson}}]{Ejiri:1999rk}%
  \BibitemOpen
  \bibfield  {author} {\bibinfo {author} {\bibfnamefont {H.}~\bibnamefont
  {Ejiri}}, \bibinfo {author} {\bibfnamefont {J.}~\bibnamefont {Engel}},
  \bibinfo {author} {\bibfnamefont {R.}~\bibnamefont {Hazama}}, \bibinfo
  {author} {\bibfnamefont {P.}~\bibnamefont {Krastev}}, \bibinfo {author}
  {\bibfnamefont {N.}~\bibnamefont {Kudomi}}, \ and\ \bibinfo {author}
  {\bibfnamefont {R.~G.~H.}\ \bibnamefont {Robertson}},\ }\bibfield  {title}
  {\enquote {\bibinfo {title} {{Spectroscopy of double beta and inverse beta
  decays from Mo-100 for neutrinos}},}\ }\href {\doibase
  10.1103/PhysRevLett.85.2917} {\bibfield  {journal} {\bibinfo  {journal}
  {Phys. Rev. Lett.}\ }\textbf {\bibinfo {volume} {85}},\ \bibinfo {pages}
  {2917--2920} (\bibinfo {year} {2000})}\BibitemShut {NoStop}%
\bibitem [{\citenamefont {Raghavan}(1997)}]{Raghavan:1997ad}%
  \BibitemOpen
  \bibfield  {author} {\bibinfo {author} {\bibfnamefont {R.~S.}\ \bibnamefont
  {Raghavan}},\ }\bibfield  {title} {\enquote {\bibinfo {title} {{New prospects
  for real time spectroscopy of low-energy electron neutrinos from the sun}},}\
  }\href {\doibase 10.1103/PhysRevLett.78.3618} {\bibfield  {journal} {\bibinfo
   {journal} {Phys. Rev. Lett.}\ }\textbf {\bibinfo {volume} {78}},\ \bibinfo
  {pages} {3618--3621} (\bibinfo {year} {1997})}\BibitemShut {NoStop}%
\bibitem [{\citenamefont {Zuber}(2003)}]{Zuber:2002wi}%
  \BibitemOpen
  \bibfield  {author} {\bibinfo {author} {\bibfnamefont {K.}~\bibnamefont
  {Zuber}},\ }\bibfield  {title} {\enquote {\bibinfo {title} {{Spectroscopy of
  low energy solar neutrinos using CdTe detectors}},}\ }\href {\doibase
  10.1016/j.physletb.2003.07.070} {\bibfield  {journal} {\bibinfo  {journal}
  {Phys. Lett.}\ }\textbf {\bibinfo {volume} {B571}},\ \bibinfo {pages}
  {148--154} (\bibinfo {year} {2003})}\BibitemShut {NoStop}%
\bibitem [{\citenamefont {Yeh}\ \emph {et~al.}(2007)\citenamefont {Yeh},
  \citenamefont {Garnov},\ and\ \citenamefont {Hahn}}]{YEH2007329}%
  \BibitemOpen
  \bibfield  {author} {\bibinfo {author} {\bibfnamefont {M.}~\bibnamefont
  {Yeh}}, \bibinfo {author} {\bibfnamefont {A.}~\bibnamefont {Garnov}}, \ and\
  \bibinfo {author} {\bibfnamefont {R.L.}\ \bibnamefont {Hahn}},\ }\bibfield
  {title} {\enquote {\bibinfo {title} {Gadolinium-loaded liquid scintillator
  for high-precision measurements of antineutrino oscillations and the mixing
  angle, $\theta_{13}$},}\ }\href {\doibase
  https://doi.org/10.1016/j.nima.2007.03.029} {\bibfield  {journal} {\bibinfo
  {journal} {Nucl. Instrum. Meth. A}\ }\textbf {\bibinfo {volume} {578}},\
  \bibinfo {pages} {329 -- 339} (\bibinfo {year} {2007})}\BibitemShut {NoStop}%
\bibitem [{\citenamefont {Graves}\ and\ \citenamefont
  {Mitchell}(1955)}]{PhysRev.97.1033}%
  \BibitemOpen
  \bibfield  {author} {\bibinfo {author} {\bibfnamefont {William~E.}\
  \bibnamefont {Graves}}\ and\ \bibinfo {author} {\bibfnamefont {Allan C.~G.}\
  \bibnamefont {Mitchell}},\ }\bibfield  {title} {\enquote {\bibinfo {title}
  {{Disintegration of ${\rm{As}}^{71}$}},}\ }\href {\doibase
  10.1103/PhysRev.97.1033} {\bibfield  {journal} {\bibinfo  {journal} {Phys.
  Rev.}\ }\textbf {\bibinfo {volume} {97}},\ \bibinfo {pages} {1033--1036}
  (\bibinfo {year} {1955})}\BibitemShut {NoStop}%
\bibitem [{\citenamefont {Morgenstern}\ \emph {et~al.}(1968)\citenamefont
  {Morgenstern}, \citenamefont {Schmidt}, \citenamefont {Fl$\ddot{u}$gge},\
  and\ \citenamefont {Schmidt}}]{MORGENSTERN1968370}%
  \BibitemOpen
  \bibfield  {author} {\bibinfo {author} {\bibfnamefont {J.}~\bibnamefont
  {Morgenstern}}, \bibinfo {author} {\bibfnamefont {J.W.}\ \bibnamefont
  {Schmidt}}, \bibinfo {author} {\bibfnamefont {G.}~\bibnamefont
  {Fl$\ddot{u}$gge}}, \ and\ \bibinfo {author} {\bibfnamefont {H.}~\bibnamefont
  {Schmidt}},\ }\bibfield  {title} {\enquote {\bibinfo {title} {The g factor of
  the 175 kev state in $^{71}\rm{Ge}$ and hyperfine fields of $^{71}\rm{Ge}$ in
  $\rm{Fe}$ and $\rm{Ni}$},}\ }\href {\doibase
  https://doi.org/10.1016/0370-2693(68)90186-X} {\bibfield  {journal} {\bibinfo
   {journal} {Physics Letters B}\ }\textbf {\bibinfo {volume} {27}},\ \bibinfo
  {pages} {370 -- 372} (\bibinfo {year} {1968})}\BibitemShut {NoStop}%
\bibitem [{\citenamefont {Taff}\ and\ \citenamefont
  {Klinken}(1978)}]{TAFF1978189}%
  \BibitemOpen
  \bibfield  {author} {\bibinfo {author} {\bibfnamefont {L.M.}\ \bibnamefont
  {Taff}}\ and\ \bibinfo {author} {\bibfnamefont {J.~Van}\ \bibnamefont
  {Klinken}},\ }\bibfield  {title} {\enquote {\bibinfo {title} {Nuclear
  halflives observed with delayed coincident summing},}\ }\href {\doibase
  https://doi.org/10.1016/0029-554X(78)90487-1} {\bibfield  {journal} {\bibinfo
   {journal} {Nuclear Instruments and Methods}\ }\textbf {\bibinfo {volume}
  {151}},\ \bibinfo {pages} {189 -- 199} (\bibinfo {year} {1978})}\BibitemShut
  {NoStop}%
\bibitem [{ens()}]{ensdf}%
  \BibitemOpen
  \href@noop {} {}\bibinfo {note} {From ENSDF database as of Nov. 18, 2019.
  Version available at http://www.nndc.bnl.gov/ensarchivals/}\BibitemShut
  {NoStop}%
\bibitem [{\citenamefont {Bahcall}(1997)}]{PhysRevC.56.3391}%
  \BibitemOpen
  \bibfield  {author} {\bibinfo {author} {\bibfnamefont {John~N.}\ \bibnamefont
  {Bahcall}},\ }\bibfield  {title} {\enquote {\bibinfo {title} {Gallium solar
  neutrino experiments: Absorption cross sections, neutrino spectra, and
  predicted event rates},}\ }\href {\doibase 10.1103/PhysRevC.56.3391}
  {\bibfield  {journal} {\bibinfo  {journal} {Phys. Rev. C}\ }\textbf {\bibinfo
  {volume} {56}},\ \bibinfo {pages} {3391--3409} (\bibinfo {year}
  {1997})}\BibitemShut {NoStop}%
\bibitem [{\citenamefont {Bahcall}(1989)}]{BahcallBook}%
  \BibitemOpen
  \bibfield  {author} {\bibinfo {author} {\bibfnamefont {John~N.}\ \bibnamefont
  {Bahcall}},\ }\href@noop {} {\emph {\bibinfo {title} {Neutrino
  Astrophysics}}}\ (\bibinfo  {publisher} {Cambridge University Press},\
  \bibinfo {year} {1989})\BibitemShut {NoStop}%
\bibitem [{\citenamefont {Bahcall}(1978)}]{RevModPhys.50.881}%
  \BibitemOpen
  \bibfield  {author} {\bibinfo {author} {\bibfnamefont {John~N.}\ \bibnamefont
  {Bahcall}},\ }\bibfield  {title} {\enquote {\bibinfo {title} {Solar neutrino
  experiments},}\ }\href {\doibase 10.1103/RevModPhys.50.881} {\bibfield
  {journal} {\bibinfo  {journal} {Rev. Mod. Phys.}\ }\textbf {\bibinfo {volume}
  {50}},\ \bibinfo {pages} {881--903} (\bibinfo {year} {1978})}\BibitemShut
  {NoStop}%
\bibitem [{\citenamefont {Giunti}\ \emph {et~al.}(2012)\citenamefont {Giunti}
  \emph {et~al.}}]{Giunti:2012tn}%
  \BibitemOpen
  \bibfield  {author} {\bibinfo {author} {\bibfnamefont {C.}~\bibnamefont
  {Giunti}} \emph {et~al.},\ }\bibfield  {title} {\enquote {\bibinfo {title}
  {{Update of Short-Baseline Electron Neutrino and Antineutrino
  Disappearance}},}\ }\href {\doibase 10.1103/PhysRevD.86.113014} {\bibfield
  {journal} {\bibinfo  {journal} {Phys. Rev.}\ }\textbf {\bibinfo {volume}
  {D86}},\ \bibinfo {pages} {113014} (\bibinfo {year} {2012})}\BibitemShut
  {NoStop}%
\bibitem [{\citenamefont {Barinov}\ \emph {et~al.}(2018)\citenamefont
  {Barinov}, \citenamefont {Cleveland}, \citenamefont {Gavrin}, \citenamefont
  {Gorbunov},\ and\ \citenamefont {Ibragimova}}]{Barinov:2017ymq}%
  \BibitemOpen
  \bibfield  {author} {\bibinfo {author} {\bibfnamefont {Vladislav}\
  \bibnamefont {Barinov}}, \bibinfo {author} {\bibfnamefont {Bruce}\
  \bibnamefont {Cleveland}}, \bibinfo {author} {\bibfnamefont {Vladimir}\
  \bibnamefont {Gavrin}}, \bibinfo {author} {\bibfnamefont {Dmitry}\
  \bibnamefont {Gorbunov}}, \ and\ \bibinfo {author} {\bibfnamefont {Tatiana}\
  \bibnamefont {Ibragimova}},\ }\bibfield  {title} {\enquote {\bibinfo {title}
  {{Revised neutrino-gallium cross section and prospects of BEST in resolving
  the Gallium anomaly}},}\ }\href {\doibase 10.1103/PhysRevD.97.073001}
  {\bibfield  {journal} {\bibinfo  {journal} {Phys. Rev.}\ }\textbf {\bibinfo
  {volume} {D97}},\ \bibinfo {pages} {073001} (\bibinfo {year}
  {2018})}\BibitemShut {NoStop}%
\bibitem [{\citenamefont {Blatt}\ and\ \citenamefont
  {Weisskopf}(1979)}]{MattWeisskopf}%
  \BibitemOpen
  \bibfield  {author} {\bibinfo {author} {\bibfnamefont {John~M.}\ \bibnamefont
  {Blatt}}\ and\ \bibinfo {author} {\bibfnamefont {Victor~F.}\ \bibnamefont
  {Weisskopf}},\ }\href@noop {} {\emph {\bibinfo {title} {Theoretical nuclear
  physics}}}\ (\bibinfo  {publisher} {Springer-Verlag},\ \bibinfo {year}
  {1979})\BibitemShut {NoStop}%
\bibitem [{\citenamefont {Frekers}\ \emph {et~al.}(2015)\citenamefont {Frekers}
  \emph {et~al.}}]{PhysRevC.91.034608}%
  \BibitemOpen
  \bibfield  {author} {\bibinfo {author} {\bibfnamefont {D.}~\bibnamefont
  {Frekers}} \emph {et~al.},\ }\bibfield  {title} {\enquote {\bibinfo {title}
  {Precision evaluation of the
  $^{71}\mathrm{Ga}({\ensuremath{\nu}}_{e},{e}^{\ensuremath{-}})$ solar
  neutrino capture rate from the $(^{3}\mathrm{He},t)$ charge-exchange
  reaction},}\ }\href {\doibase 10.1103/PhysRevC.91.034608} {\bibfield
  {journal} {\bibinfo  {journal} {Phys. Rev. C}\ }\textbf {\bibinfo {volume}
  {91}},\ \bibinfo {pages} {034608} (\bibinfo {year} {2015})}\BibitemShut
  {NoStop}%
\bibitem [{\citenamefont {Haxton}(1998)}]{Haxton:1998uc}%
  \BibitemOpen
  \bibfield  {author} {\bibinfo {author} {\bibfnamefont {W.~C.}\ \bibnamefont
  {Haxton}},\ }\bibfield  {title} {\enquote {\bibinfo {title} {{Cross-section
  uncertainties in the gallium neutrino source experiments}},}\ }\href
  {\doibase 10.1016/S0370-2693(98)00581-4} {\bibfield  {journal} {\bibinfo
  {journal} {Phys. Lett.}\ }\textbf {\bibinfo {volume} {B431}},\ \bibinfo
  {pages} {110--118} (\bibinfo {year} {1998})}\BibitemShut {NoStop}%
\bibitem [{\citenamefont {Biller}(2015)}]{BILLER2015205}%
  \BibitemOpen
  \bibfield  {author} {\bibinfo {author} {\bibfnamefont {Steven}\ \bibnamefont
  {Biller}},\ }\bibfield  {title} {\enquote {\bibinfo {title} {$\mathrm{SNO}$+
  with tellurium},}\ }\href {\doibase
  https://doi.org/10.1016/j.phpro.2014.12.033} {\bibfield  {journal} {\bibinfo
  {journal} {Physics Procedia}\ }\textbf {\bibinfo {volume} {61}},\ \bibinfo
  {pages} {205 -- 210} (\bibinfo {year} {2015})},\ \bibinfo {note} {13th
  International Conference on Topics in Astroparticle and Underground Physics,
  TAUP 2013}\BibitemShut {NoStop}%
\bibitem [{\citenamefont {Raghavan}\ and\ \citenamefont {the
  LENS~Collaboration}(2008)}]{Raghavan_2008}%
  \BibitemOpen
  \bibfield  {author} {\bibinfo {author} {\bibfnamefont {R~S}\ \bibnamefont
  {Raghavan}}\ and\ \bibinfo {author} {\bibnamefont {the LENS~Collaboration}},\
  }\bibfield  {title} {\enquote {\bibinfo {title} {{LENS},
  {MiniLENS}{\textemdash}status and outlook},}\ }\href {\doibase
  10.1088/1742-6596/120/5/052014} {\bibfield  {journal} {\bibinfo  {journal}
  {Journal of Physics: Conference Series}\ }\textbf {\bibinfo {volume} {120}},\
  \bibinfo {pages} {052014} (\bibinfo {year} {2008})}\BibitemShut {NoStop}%
\bibitem [{\citenamefont {Ashenfelter}\ \emph {et~al.}(2018)\citenamefont
  {Ashenfelter} \emph {et~al.}}]{Ashenfelter:2018cli}%
  \BibitemOpen
  \bibfield  {author} {\bibinfo {author} {\bibfnamefont {J.}~\bibnamefont
  {Ashenfelter}} \emph {et~al.} (\bibinfo {collaboration} {PROSPECT}),\
  }\bibfield  {title} {\enquote {\bibinfo {title} {{Performance of a segmented
  $^{6}$Li-loaded liquid scintillator detector for the PROSPECT experiment}},}\
  }\href {\doibase 10.1088/1748-0221/13/06/P06023} {\bibfield  {journal}
  {\bibinfo  {journal} {JINST}\ }\textbf {\bibinfo {volume} {13}},\ \bibinfo
  {pages} {P06023} (\bibinfo {year} {2018})}\BibitemShut {NoStop}%
\bibitem [{\citenamefont {Yeh}\ \emph {et~al.}(2011)\citenamefont {Yeh} \emph
  {et~al.}}]{YEH201151}%
  \BibitemOpen
  \bibfield  {author} {\bibinfo {author} {\bibfnamefont {M.}~\bibnamefont
  {Yeh}} \emph {et~al.},\ }\bibfield  {title} {\enquote {\bibinfo {title} {A
  new water-based liquid scintillator and potential applications},}\ }\href
  {\doibase https://doi.org/10.1016/j.nima.2011.08.040} {\bibfield  {journal}
  {\bibinfo  {journal} {Nucl. Instrum. Meth. A}\ }\textbf {\bibinfo {volume}
  {660}},\ \bibinfo {pages} {51 -- 56} (\bibinfo {year} {2011})}\BibitemShut
  {NoStop}%
\bibitem [{Ham()}]{Hamamatsu}%
  \BibitemOpen
  \href@noop {} {}\bibinfo {note} {Hamamatsu Photonics, see also the web link
  http://www.hamamatsu.com/}\BibitemShut {NoStop}%
\bibitem [{cae()}]{caen}%
  \BibitemOpen
  \href@noop {} {}\bibinfo {note} {CAEN, see also the web link
  https://www.caen.it/}\BibitemShut {NoStop}%
\bibitem [{\citenamefont {Aharmim}\ \emph {et~al.}(2005)\citenamefont {Aharmim}
  \emph {et~al.}}]{Aharmim:2005gt}%
  \BibitemOpen
  \bibfield  {author} {\bibinfo {author} {\bibfnamefont {B.}~\bibnamefont
  {Aharmim}} \emph {et~al.} (\bibinfo {collaboration} {SNO}),\ }\bibfield
  {title} {\enquote {\bibinfo {title} {{Electron energy spectra, fluxes, and
  day-night asymmetries of B-8 solar neutrinos from measurements with NaCl
  dissolved in the heavy-water detector at the Sudbury Neutrino
  Observatory}},}\ }\href {\doibase 10.1103/PhysRevC.72.055502} {\bibfield
  {journal} {\bibinfo  {journal} {Phys. Rev.}\ }\textbf {\bibinfo {volume}
  {C72}},\ \bibinfo {pages} {055502} (\bibinfo {year} {2005})}\BibitemShut
  {NoStop}%
\bibitem [{\citenamefont {Aharmim}\ \emph {et~al.}(2007)\citenamefont {Aharmim}
  \emph {et~al.}}]{Aharmim:2006kv}%
  \BibitemOpen
  \bibfield  {author} {\bibinfo {author} {\bibfnamefont {B.}~\bibnamefont
  {Aharmim}} \emph {et~al.} (\bibinfo {collaboration} {SNO}),\ }\bibfield
  {title} {\enquote {\bibinfo {title} {{Determination of the $\nu_e$ and total
  $^8$B solar neutrino fluxes with the Sudbury neutrino observatory phase I
  data set}},}\ }\href {\doibase 10.1103/PhysRevC.75.045502} {\bibfield
  {journal} {\bibinfo  {journal} {Phys. Rev.}\ }\textbf {\bibinfo {volume}
  {C75}},\ \bibinfo {pages} {045502} (\bibinfo {year} {2007})}\BibitemShut
  {NoStop}%
\bibitem [{\citenamefont {Serenelli}\ \emph
  {et~al.}(2009{\natexlab{b}})\citenamefont {Serenelli}, \citenamefont {Basu},
  \citenamefont {Ferguson},\ and\ \citenamefont {Asplund}}]{AGS09}%
  \BibitemOpen
  \bibfield  {author} {\bibinfo {author} {\bibfnamefont {Aldo}\ \bibnamefont
  {Serenelli}}, \bibinfo {author} {\bibfnamefont {Sarbani}\ \bibnamefont
  {Basu}}, \bibinfo {author} {\bibfnamefont {Jason~W.}\ \bibnamefont
  {Ferguson}}, \ and\ \bibinfo {author} {\bibfnamefont {Martin}\ \bibnamefont
  {Asplund}},\ }\bibfield  {title} {\enquote {\bibinfo {title} {{New Solar
  Composition: The Problem With Solar Models Revisited}},}\ }\href {\doibase
  10.1088/0004-637X/705/2/L123} {\bibfield  {journal} {\bibinfo  {journal}
  {Astrophys. J. Lett.}\ }\textbf {\bibinfo {volume} {705}},\ \bibinfo {pages}
  {L123} (\bibinfo {year} {2009}{\natexlab{b}})}\BibitemShut {NoStop}%
\bibitem [{\citenamefont {Tanabashi}\ \emph {et~al.}(2018)\citenamefont
  {Tanabashi} \emph {et~al.}}]{PDG2018}%
  \BibitemOpen
  \bibfield  {author} {\bibinfo {author} {\bibfnamefont {M.}~\bibnamefont
  {Tanabashi}} \emph {et~al.} (\bibinfo {collaboration} {Particle Data
  Group}),\ }\bibfield  {title} {\enquote {\bibinfo {title} {{Review of
  Particle Physics}},}\ }\href {\doibase 10.1103/PhysRevD.98.030001} {\bibfield
   {journal} {\bibinfo  {journal} {Phys. Rev.}\ }\textbf {\bibinfo {volume}
  {D98}},\ \bibinfo {pages} {030001} (\bibinfo {year} {2018})}\BibitemShut
  {NoStop}%
\bibitem [{\citenamefont {Beacom}\ \emph {et~al.}(2017)\citenamefont {Beacom}
  \emph {et~al.}}]{Jinping}%
  \BibitemOpen
  \bibfield  {author} {\bibinfo {author} {\bibfnamefont {John~F.}\ \bibnamefont
  {Beacom}} \emph {et~al.} (\bibinfo {collaboration} {Jinping}),\ }\bibfield
  {title} {\enquote {\bibinfo {title} {{Physics prospects of the Jinping
  neutrino experiment}},}\ }\href {\doibase 10.1088/1674-1137/41/2/023002}
  {\bibfield  {journal} {\bibinfo  {journal} {Chin. Phys. C}\ }\textbf
  {\bibinfo {volume} {41}},\ \bibinfo {pages} {023002} (\bibinfo {year}
  {2017})},\ \Eprint {http://arxiv.org/abs/1602.01733} {arXiv:1602.01733
  [physics.ins-det]} \BibitemShut {NoStop}%
\bibitem [{\citenamefont {Bolozdynya}\ \emph {et~al.}(2012)\citenamefont
  {Bolozdynya} \emph {et~al.}}]{Bolozdynya:2012xv}%
  \BibitemOpen
  \bibfield  {author} {\bibinfo {author} {\bibfnamefont {A.}~\bibnamefont
  {Bolozdynya}} \emph {et~al.},\ }\bibfield  {title} {\enquote {\bibinfo
  {title} {{Opportunities for Neutrino Physics at the Spallation Neutron
  Source: A White Paper}},}\ \ }(\bibinfo {year} {2012})\ \Eprint
  {http://arxiv.org/abs/1211.5199} {arXiv:1211.5199 [hep-ex]} \BibitemShut
  {NoStop}%
\bibitem [{\citenamefont {Chen}\ and\ \citenamefont {Wang}(2016)}]{CSNS}%
  \BibitemOpen
  \bibfield  {author} {\bibinfo {author} {\bibfnamefont {H.}~\bibnamefont
  {Chen}}\ and\ \bibinfo {author} {\bibfnamefont {X.}~\bibnamefont {Wang}},\
  }\bibfield  {title} {\enquote {\bibinfo {title} {{China's first pulsed
  neutron source}},}\ }\href {\doibase doi:10.1038/nmat4655} {\bibfield
  {journal} {\bibinfo  {journal} {Nature Materials}\ }\textbf {\bibinfo
  {volume} {15}},\ \bibinfo {pages} {689--691} (\bibinfo {year}
  {2016})}\BibitemShut {NoStop}%
\end{thebibliography}%

\end{document}